\documentclass[aps,prd,amssymb,amsmath,amsfonts,superscriptaddress,nofootinbib,reprint,showpacs,longbibliography]{revtex4-2}

\usepackage{graphicx}
\usepackage{lmodern}
\usepackage{amsmath,amssymb}
\usepackage{mathrsfs}
\usepackage{amsfonts}
\usepackage[utf8]{inputenc}
\usepackage{url}
\usepackage[colorlinks]{hyperref}
\usepackage{xcolor}
\usepackage[normalem]{ulem}
\usepackage{orcidlink}
\usepackage{placeins}
\usepackage{mathtools}
\usepackage{rotating}
\usepackage{tabularx}
\usepackage{enumitem} 
\usepackage{xspace}
\usepackage{subcaption}
\usepackage{multirow}
\usepackage{booktabs}
\usepackage[justification=justified,singlelinecheck=false]{caption}

\begin{document}

\title{Extending multi-messenger constraints on neutron star matter through the inclusion of direct Urca cooling}

\author{M Dio \surname{Danarianto}\orcidlink{0000-0001-9801-7684}}
\affiliation{ Institut f\"ur Physik und Astronomie, Universit\"at Potsdam, Haus 28, Karl-Liebknecht-Str. 24/25, 14476, Potsdam, Germany}
\affiliation{ Research Center for Computing, National Research and Innovation Agency (BRIN), Bandung, 40173, Indonesia}
\author{Guilherme \surname{Grams}\orcidlink{0000-0002-8635-383X}}
\affiliation{ Institut f\"ur Physik und Astronomie, Universit\"at Potsdam, Haus 28, Karl-Liebknecht-Str. 24/25, 14476, Potsdam, Germany}
\author{Thibeau~Wouters\orcidlink{0009-0006-2797-3808}}
{\affiliation{Institute for Gravitational and Subatomic Physics (GRASP), Utrecht University, Princetonplein 1, 3584 CC Utrecht, The Netherlands}
\affiliation{Nikhef, Science Park 105, 1098 XG Amsterdam, The Netherlands}
\author{Tim \surname{Dietrich}\orcidlink{0000-0003-2374-307X}}
\affiliation{ Institut f\"ur Physik und Astronomie, Universit\"at Potsdam, Haus 28, Karl-Liebknecht-Str. 24/25, 14476, Potsdam, Germany}
\affiliation{ Max Planck Institute for Gravitational Physics (Albert Einstein Institute), Am M\"uhlenberg 1, Potsdam 14476, Germany}

\date{\today}
             
\begin{abstract}
The properties and behavior of strongly interacting matter at extreme densities can not only be determined through terrestrial experiments or through theoretical considerations, but also astrophysical observations of neutron stars (NSs) and binary neutron star systems become increasingly important to obtain complementary information. In light of these findings, we perform a Bayesian study inferring the NS equation of state (EoS) by incorporating constraints from the direct Urca (dUrca) process, thereby including information on matter composition in addition to traditionally used macroscopic observables. 
We implement a self-consistent treatment of $\beta$-equilibrium and charge neutrality in the \textsc{jester} framework, including the calculation of the proton fraction and the onset of nucleonic dUrca in the presence of electrons and muons. 
We combine constraints from chiral effective field theory, astrophysical measurements of NS masses, radii, and tidal deformabilities, and observations of rapidly and slowly cooling neutron stars. 
The dUrca constraint seems to be in favor of stiffer EoSs, but --at the current stage-- has only a minor impact on macroscopic NS properties once we include other nuclear and astrophysical constraints. In contrast, the dUrca information provides a constraint on the composition of canonical-mass NSs favoring a higher proton fraction inside the star, illustrating the complementary information carried by cooling observations. 
Our results demonstrate the potential of incorporating composition-sensitive observables into multimessenger inference and provide a step toward a more comprehensive treatment of dUrca constraints with microscopically motivated EoS models.
\end{abstract}

\maketitle

\section{Introduction} Despite noticeable progress over the last years, cf.~e.g.~\cite{Lattimer:2012nd,Burgio:2021vgk,Chatziioannou:2024jsr,Koehn:2024set} for reviews and references, understanding the properties of strongly interacting matter at the highest densities in our Universe is still an unanswered question of modern physics. Although nuclear physics computations can provide important information up to about two times nuclear saturation density ($2\ n_{\rm sat}$), i.e., twice the densities reached typically inside atomic nuclei, through methods such as chiral effective field theory~\cite{Machleidt:2011zz,Drischler:2021kxf}, or at extremely high densities ($\gtrsim 40 n_{\rm sat}$), where quantum chromodynamics can be solved perturbatively~\cite{Kurkela:2009gj,Komoltsev:2021jzg}, the intermediate regime of a few times nuclear saturation density remains theoretically uncertain and is difficult to probe directly in terrestrial experiments~\cite{Burgio:2021vgk,Drischler:2021kxf,Tsang:2024tqg}.
Neutron stars (NSs), among the densest objects known in our Universe, offer a unique opportunity to probe this density region and provide important complementary information.
Luckily, observations of NSs or NS mergers can happen through numerous observational channels, which --when combined-- provide a wealth of information that can be used to constrain the properties of strongly interacting matter.
Such observations include radio observations of massive pulsars~\cite{Antoniadis:2013pzd,NANOGrav:2017wvv,Rezzolla:2017aly} revealing the maximum mass of NSs, X-ray pulse-profile modelling of rotating NSs through facilities such as NICER providing mass-radius measurements~\cite{Riley:2019yda,Miller:2019cac,Riley:2021pdl,Miller:2021qha,Vinciguerra2024,Choudhury2024,Salmi2024,Mauviard2025,Kini:2026rjx,Miller:2025qfq,Miller:2026vpr}, as well as multimessenger observations of merging binary NSs, for which GW170817/GRB170817A/AT2017gfo~\cite{LIGOScientific:2017vwq,Cowperthwaite_2017,LIGOScientific:2018hze,LIGOScientific:2018cki} is an exemplary case. We refer further to~\cite{Koehn:2024set} for a review of many of the existing observational constraints.

The direct Urca (dUrca) process, consisting of neutron beta decay and its inverse electron-capture reaction, can occur only when the proton fraction exceeds a threshold value. For npe matter, i.e., matter consisting of neutrons, protons, and electrons, this threshold is $\simeq 11.1\%$, increasing to $\sim14.8\%$ when muons are present~\cite{Klahn06}. Since the proton fraction in $\beta$-equilibrated NS matter is governed by the density dependence of the nuclear symmetry energy, the dUrca threshold directly connects NS cooling to the composition and nuclear EoS~\cite{Margueron:2017lup,Burgio:2021vgk}. Importantly, rapid cooling is not unique to nucleonic dUrca: fast neutrino emission may also arise from exotic degrees of freedom such as hyperons or deconfined quarks, while nucleonic dUrca can be substantially suppressed by nucleon superfluidity~\cite{Yakovlev:2000jp,Burgio:2021vgk,Yakovlev:2004iq,Potekhin:2015qsa}.
However, superfluidity itself could also accelerate cooling, which makes the identification of a rapidly cooling source with nucleonic dUrca is not unique~\cite{Page:2010aw,Shternin:2010qi}.

Several studies have investigated the relation between dUrca, the nuclear symmetry energy, and NS properties, including Bayesian analyses based on nucleonic metamodels and nuclear energy-density functionals~\cite{Margueron:2017lup,Malik:2022yaj,Beznogov:2015qra,Beznogov:2024jle}. 
Very recently, Ref.~\cite{Montefusco:2026dU} combined nuclear-structure and isospin-transport data with $\chi$EFT and astrophysical constraints within an asymptotically causal metamodel~\cite{Montefusco:2026jlq}, finding that nucleonic dUrca onset at or below $1.4\,M_\odot$ is strongly disfavored.
Recent multimessenger EoS analyses have also used nuclear and astronomical constraints to infer the proton fraction and assess Urca cooling~\cite{Tsang:2024tqg}. 
Here, we take a complementary step by incorporating the dUrca information explicitly into the likelihood of a multimessenger Bayesian EoS inference. In contrast to macroscopic observables such as masses, radii, and tidal deformabilities, evaluating dUrca requires a model that provides the composition of $\beta$-equilibrated matter. We therefore extend the \textsc{jester} framework~\cite{Wouters:2025zju} to consistently compute the proton fraction and dUrca threshold within different nucleonic models, while retaining an agnostic high-density sound-speed extension. This allows us to quantify how dUrca information affects both the inferred macroscopic properties of NSs and the proton fraction of canonical-mass stars.

\section{EOS construction} 

We extend the available EoS classes in the \textsc{jester} framework and we use a metamodel~\cite{Margueron:2017eqc} with a sound-speed extension~\cite{Tews:2018iwm,Dietrich:2020efo,Koehn:2024set} and also include the extended-Skyrme energy-density functionals~\cite{Chamel:2009yx,Grams:2023sml}. 
In the nucleonic model, we further incorporate muons in the lepton sector and solve $\beta$-equilibrium self-consistently,\footnote{Previous implementations in \textsc{jester} used the approximation 
$\mu_e = \mu_n - \mu_p \simeq 4E_{\rm sym}(n)\,(1-2Y_p),$ 
to estimate the proton fraction of NS matter. } 
and compute direct Urca thresholds for both electron and muon channels. These enable a more complete and microphysically consistent assessment of the impact of dUrca.

In this work, we keep the NS crust fixed to the DH(SLy4) EoS \cite{Douchin:2001sv} and use a metamodel and extended-Skyrme EoS for the homogeneous nuclear matter. We consider two assumptions about the nucleonic EoS breakdown density. 
First, we use a variable breakdown density, where we sample $n_{\rm break}$ randomly between $1~n_{\rm sat}$ to $4~n_{\rm sat}$. This captures a wide range of EoSs on the speed-of-sound branch and is a widely adopted assumption \cite{Koehn:2024set}. 
Second, we determine $n_{\rm break}$ as the density when the speed of sound within the nucleonic model reaches either zero or one, i.e., when the model breaks down. This provides an upper limit on $ n_{\rm break}$, enabling a wider parameter space of the nucleonic EoS. 

\begin{figure}
    \centering
    \includegraphics[width=1\linewidth]{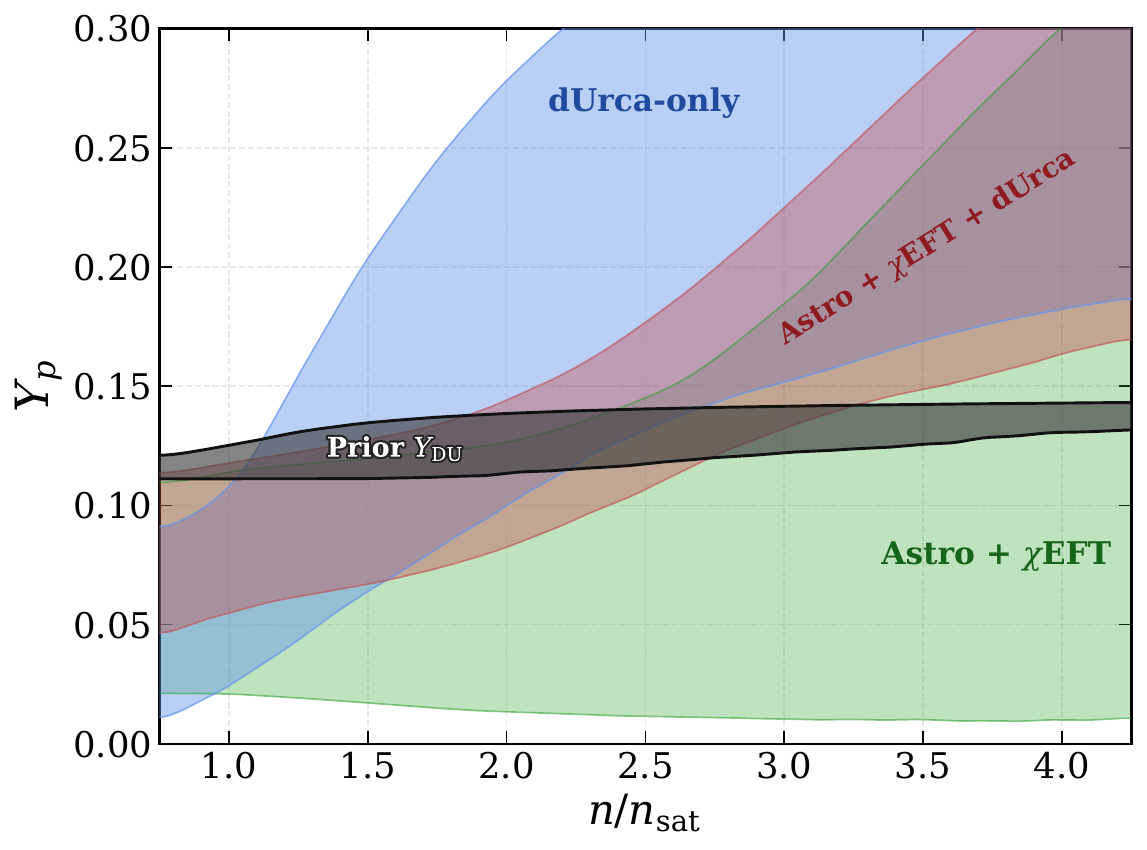}
    \caption{Proton fraction $Y_p$ versus baryon density $n/n_{\rm sat}$ of extended-Skyrme EOS. The gray shaded band denotes the 95\% prior region for the direct Urca threshold $Y_{\rm DU}$.
    The shaded regions show the 95\% credible interval of posteriors using dUrca only (blue), Astro + $\chi$EFT (green),  and Astro + $\chi$EFT + dUrca (red). Details about the individual Astro and $\chi$EFT constraints are outlined in Sec.~\ref{sec:Bayesian}.}
    \label{fig:ypxdu}
\end{figure}

\begin{figure}
    \centering
    \includegraphics[width=1\linewidth]{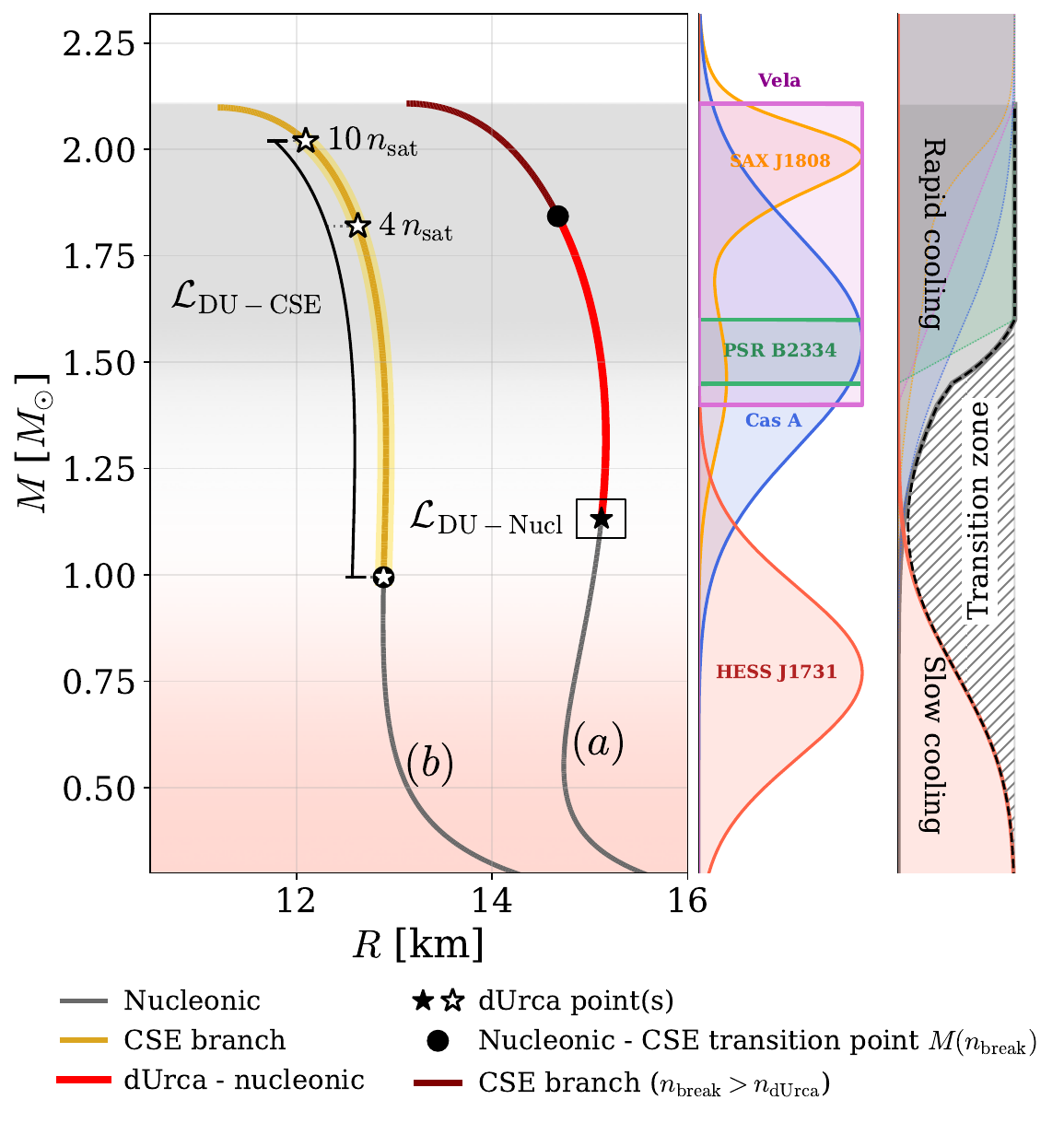}
    \caption{Schematic of the dUrca likelihood evaluation on mass–radius curves for representative EOS construction. Left panel: Evaluated mass–radius branches under distinct onset regimes. Curve ($a$) shows nucleonic dUrca onset at $n_{\rm DU} < n_{\rm break}$ (black star), with NSs satisfying $n_c \ge n_{\rm DU}$ highlighted in red, followed by the CSE extension past $n_{\rm break}$ (black circle, dark red). Curve ($b$) shows models without nucleonic dUrca ($n_{\rm DU} > n_{\rm break}$ or absent); the yellow CSE segment indicates where the dUrca likelihood is marginalized from $M(n_{\rm break})$ (star-in-circle) to $M(n^*)$ across $n^* \in [4, 10]\,n_{\rm sat}$ (white stars). Background shading showing the joint dUrca likelihood, with lighter color marking higher probability. Middle panel: Mass probability distributions for rapidly cooling sources \cite{Heinke:2006ie,Dorsman:2025bzc,Ho:2009mm, Ho:2021hwy, Shternin:2022rti, McGowan:2005kt, Vigano:2013lea, Beloin:2018fyp, Negi:2005ip} and the slow-cooling source HESS J1731\textminus347 \cite{Klochkov:2014ola,Doroshenko:2022nwp}. Right panel: Cumulative distribution functions showing joint exclusion limits based on slow-cooling and rapid-cooling regimes, with transition zone defining the likelihood of dUrca mass $M(n_{\rm DU})$ point.}
    \label{fig:mrcurve_and_bounds}
\end{figure}

Generally, observations suggest that at sufficiently high densities and above a given proton fraction, $Y_p \ge Y_{\rm DU}$, dUrca can set in. 
$Y_{\rm DU}$ is determined by enforcing $\beta$-equilibrium, charge-neutrality for zero-temperature hadronic matter consisting of protons $(p)$, neutrons $(n)$, electrons $(e)$, and muons $(\mu)$.

Enforcing $\beta$-equilibrium and momentum conservation, the Fermi surfaces of the particles fulfill 
$    p_{F,n} \le p_{F,p} + p_{F,e},$
which allows for the direct Urca $n \rightarrow p + e^- + \bar{\nu}_e$ and its inverse reaction to occur. Using the relation 
$p_{F,i} = \hbar(3\pi^2n_i)^{1/3},$ with particle fractions defined by $Y_i = n_i/n$, this condition equates to:
$(1-Y_p)^{1/3} \le Y_p^{1/3} + Y_e^{1/3},$
which can be simplified into:
\begin{equation}
    Y_p \ge \left(1+\left(1+\left({Y_e}/{Y_p}\right)^{1/3}\right)^3\right)^{-1} = Y_{\rm DU},
\end{equation}
with $Y_p=Y_e + Y_{\mu}$, where $Y_{\rm DU}$ defines the onset of the dUrca process \cite{Lattimer:1991ib,Klahn06}. 

If this condition is fulfilled in the center of the star, thermal energy will be quickly converted into neutrino radiation that escapes the NS, hence, the temperature will decrease rapidly. Therefore, for a certain EoS, there might be a density $n_{\rm DU}$ at which the proton fraction exceeds $Y_{\rm DU}$ and the direct Urca process sets in. Figure~\ref{fig:ypxdu} shows $Y_{\rm DU}$ together with the prior proton fraction for various setups as further discussed throughout this paper. 

We can have the following scenarios. In the simplest case, dUrca sets in within the nucleonic part of the EOS described by the nucleonic model; see label (a) in Fig.~\ref{fig:mrcurve_and_bounds}. In this case, we can directly compute the NS mass at which dUrca sets in and compare to observational constraints. 
Another possibility is that the nucleonic description terminates at low densities, either because the randomly chosen breakdown density $n_{\rm break}$ is reached before the dUrca density, or because the sampled metamodel parametrization becomes unphysical (e.g., reaches its causality limit). In these cases, the nucleonic branch does not extend to densities high enough to determine whether dUrca took place; see label (b) in Fig.~\ref{fig:mrcurve_and_bounds}.
Hence, when the density exceeds $n_{\rm break}$, we use a sound-speed extension (CSE). In this work, we use a six-segment extension up to $25n_{\rm sat}$. This provides greater flexibility, covers a wider range of stiffness, and reaches massive NSs in an agnostic fashion. However, no microscopic description can be deduced within the CSE region. 

\section{Bayesian inference setup and observational constraints} 
\label{sec:Bayesian}
In addition to our EOS construction, the right panels of Fig.~\ref{fig:mrcurve_and_bounds} indicate the measured masses of objects whose temperature and age place them below the predictions of the minimal cooling paradigm~\cite{Page:2004fy}, i.e., objects inferred to be rapidly cooling, for which we therefore assume dUrca is active. These objects are SAX J1808.4\textminus3658 \cite{Heinke:2006ie,Dorsman:2025bzc} for which the cooling profile and the mass are independently measured. 
On the other hand, we have Cassiopeia A \cite{Ho:2009mm, Ho:2021hwy, Shternin:2022rti}, PSR B2334+61 \cite{McGowan:2005kt}, and the Vela pulsar (PSR B0833\textminus45) \cite{Vigano:2013lea, Beloin:2018fyp, Negi:2005ip}, for which their masses have been derived from their cooling profile or spectra. 
At lower masses, the temperature of a compact object is high enough to indicate the absence of rapid cooling, so we assume the dUrca process is not occurring in this region. We use observations of XMMU J173203.3-344518~\cite{Klochkov:2014ola} central compact object associated with HESS J1731-347 whose mass is independently measured. We emphasize that the mass and radius of HESS J1731-347 measured in~\cite{Doroshenko:2022nwp} is strongly under debate, as using more recent observational data or distance measurements noticeably shifts the mass and radius towards larger values. However, we incorporate this as a relaxed constraint here as a proof of principle on how observations of non-dUrca for low mass NSs can be incorporated in the framework. 

Using the TOV equations~\cite{Tolman1939,Oppenheimer1939}, i.e., working under the assumption that the considered stars are stationary, spherically symmetric and can be described by an ideal fluid in general relativity, a given central density $n_c$ can be mapped to a given NS mass. This enables us to use mass estimates of both rapidly (labeled as data A) and slowly cooling objects (labeled as data B) to determine $n_{\rm dUrca}$. 
Following this approach, the transition region (Fig.~\ref{fig:mrcurve_and_bounds}) is computed following
\begin{equation}
\begin{split}
\mathcal{L}_\text{DU, trans} =  \qquad & \\
\underbrace{\prod_A (1 - \text{CDF}(M_{\text{(A)}}))}_{\rm upper\ limit} & 
\times 
\underbrace{\prod_B \text{CDF}(M_{\text{(B)}})}_{\rm lower\ limit},
\end{split}
\label{eq:likelihooddurca}
\end{equation} 
where  $M_{\text{(A,B)}}$  describe the masses of the rapidly or slowly cooling objects, respectively. The onset of dUrca ($n_c \ge n_{\rm DU}$) corresponds to a threshold mass $M_{\rm DU}$ within the `Transition zone' $\mathcal{L}_\text{DU, trans}$.

Given that the transition can happen within the nucleonic part of the EOS or within the agnostic extension, the overall likelihood of the dUrca constraint has to be split:
\begin{equation*}
	P(\text{data} \mid M_{\text{DU}}) = 
	\begin{cases} 
		\mathcal{L}_{\text{DU-Nucl}}, & \text{if } n_{\text{DU}} < n_{\text{break}}, \\
		\mathcal{L}_{\text{DU-CSE}}, & \text{otherwise}.
	\end{cases}
\end{equation*}
The first case is simply the likelihood of the transition zone, 
\begin{equation}
\mathcal{L}_{\text{DU-Nucl}} = \mathcal{L}_{\rm DU, trans}(n_{\rm DU},{\rm EOS}_i).
\end{equation} 
Since it is not possible to compute the proton fraction within the CSE branch, we use the marginalized probability
\begin{equation}
\begin{split}
	\mathcal{L}_{\text{DU-CSE}} = \int & \mathcal{L}_{\rm DU, trans}(n_{\rm DU},{\rm EOS}_i) \\ & \pi(n_{\rm DU}) \pi(n^*) \, dn_{\rm DU} ~dn^*,
    \end{split}
\end{equation}
where $n_\text{DU}$ represents the density at which the dUrca sets in. For our analysis, we use a uniform prior on $n_\text{DU}$, $\pi (n_{\rm DU}) = \mathcal{U}(n_{\text{break}}, n^*)$ with $n^* \in \pi(n^*) = \mathcal{U}(4 n_{\rm sat}, 10 n_{\rm sat})$. When $n_{\rm TOV}<n_{\rm DU}$ and/or $n_{\rm break}>4 n_{\rm sat}$, the probability is set to zero. 

\begin{figure*}[t!]
    \centering
    \includegraphics[width=0.99\textwidth]{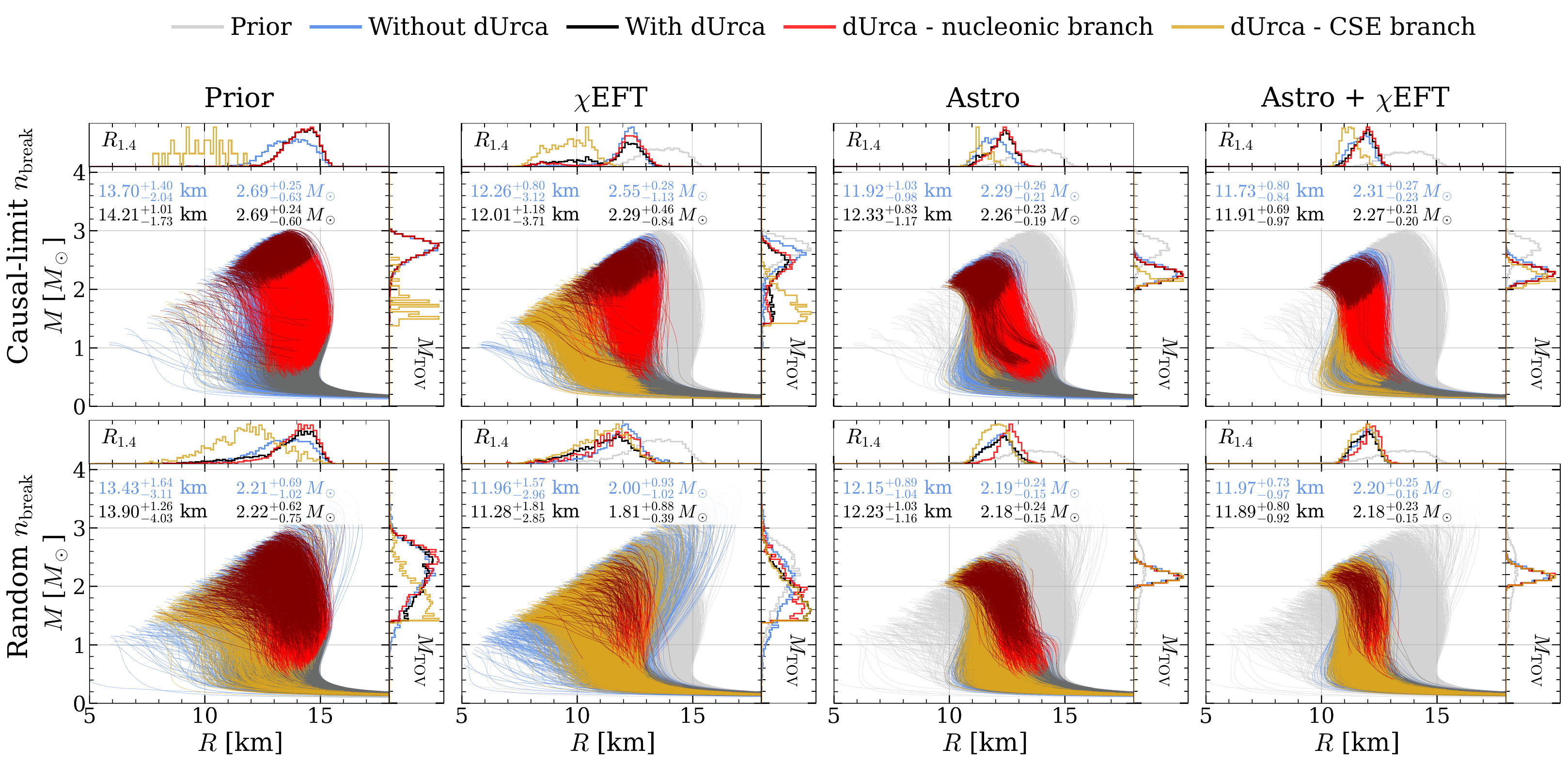} 
    \caption{Mass-radius posterior predictive for causal (top) and random (bottom) $n_{\rm break}$ with metamodel EoS. Blue lines showing posterior samples with no cooling constraints. Samples with dUrca constraints are shown in red if the dUrca sets in inside the nucleonic branch and yellow if it sets in the CSE branch. Dark red refers to the continuation of the nucleonic branch (red). Upper subplot shows $R_{1.4}$ histograms and right subplot shows $M_{\rm TOV}$. Black lines in the histograms shows overall cooling constraints. Light gray lines showing prior distribution as comparison while dark gray lines at the lower masses showing values below dUrca density for dUrca branch or $n_{\rm break}$ for CSE branch.}\label{fig:Combined_Random_nbreak_pub_mass_radius_hist}
\end{figure*}
\begin{figure*}[t!]
    \centering
    \includegraphics[width=0.99\textwidth]{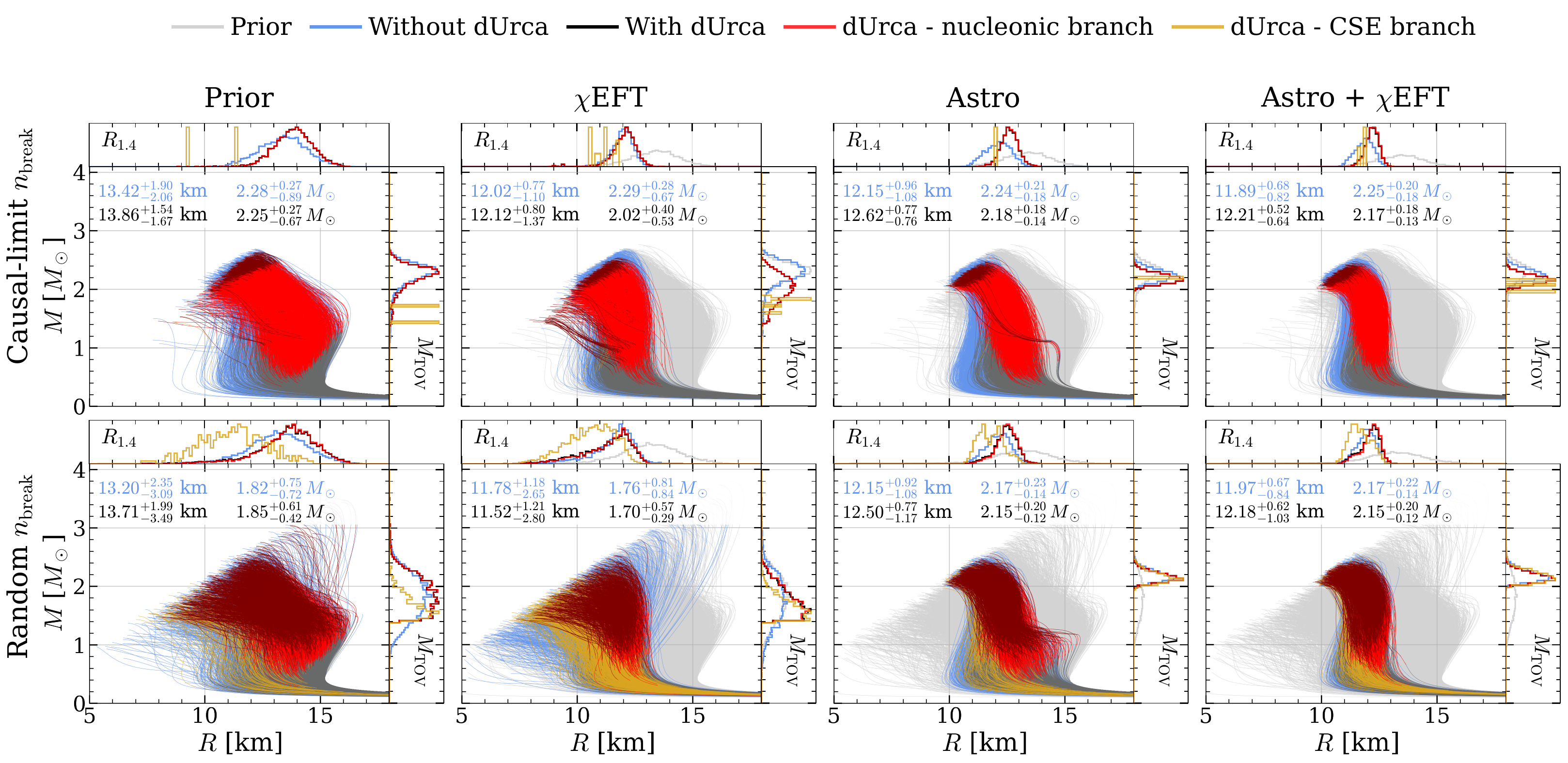} 
    \caption{Same as Fig.~\ref{fig:Combined_Random_nbreak_pub_mass_radius_hist} but for extended-Skyrme.} 
\label{fig:Combined_Random_nbreak_pub_mass_radius_hist_Skyrme}
\end{figure*}

Beyond the dUrca constraint, we combine our analysis with $\chi\text{EFT}$ \cite{Koehn:2024set} alongside multiple astrophysical observations. The astrophysical constraints incorporate tidal deformability measurements from gravitational-wave event GW170817 under the GWTC\textminus1 low-spin posterior \cite{LIGOScientific:2017vwq,LIGOScientific:2018hze}, joint mass-radius estimates of PSR J0030+0451 \cite{Riley:2019yda,Miller:2019cac}, PSR J0740+6620 \cite{Salmi2024,Miller:2021qha}, PSR J0437\textminus4715 \cite{Choudhury2024,Miller:2025qfq}, and PSR J0614\textminus3329 \cite{Mauviard2025}, heavy radio pulsar mass measurements of PSR J1614\textminus2230 \cite{NANOGrav:2017wvv} and PSR J0348+0432 \cite{Antoniadis:2013pzd}, and the maximum mass upper limit from \cite{Rezzolla:2017aly}. We construct the joint astrophysical likelihood as $\mathcal{L}_{\text{astro}} = \mathcal{L}_{\Lambda} \mathcal{L}_{MR} \mathcal{L}_{M_{\text{TOV}}}$, where $\mathcal{L}_{\Lambda}$ and $\mathcal{L}_{MR}$ are evaluated from posterior combinations and mixture distributions taken from the referenced sources, while $\mathcal{L}_{M_{\text{TOV}}}$ follows the formulation in \cite{Dietrich:2020efo}. The total likelihood is defined as the product across individual contributions, $\mathcal{L}_{\text{total}} = \prod_{i} \mathcal{L}_{i}$ with $\mathcal{L}_{i} \in \{\mathcal{L}_{\text{DU}}, \mathcal{L}_{\text{astro}}, \mathcal{L}_{\chi\text{EFT}}\}$, and each combination of constraints is evaluated both with and without the direct Urca condition.

Employing the Bayesian setup described above, we use the \textsc{jester} framework~\cite{Wouters:2025zju} to obtain our results. \textsc{jester} utilizes a JAX-based backend to enable native GPU acceleration~\cite{frostig2018compiling,cabezas2024blackjax}. Posterior distributions are sampled using a sequential Monte Carlo algorithm configured with $4000$ particles~\cite{del2006sequential,chopin2020introduction,dai2022invitation}.\footnote{To be specific, using an NVIDIA Quadro RTX 6000 GPU (24GB VRAM), a typical inference run requires an average execution time of approximately 2 hours per case, varying by about 1 hour depending on the number of constraints evaluated simultaneously.}

\section{Results and discussions} 

We show the mass-radius diagrams for the metamodel within the causal (random) $n_{\rm break}$ construction in top (bottom) panels of Fig.~\ref{fig:Combined_Random_nbreak_pub_mass_radius_hist}. From left to right, we display the prior, the inclusion of constraints from $\chi$EFT, Astro, and combined $\chi$EFT + Astro.
Light red curves show NS where the dUrca threshold was achieved by the nucleonic EoS, with dark red showing its CSE extension.
Yellow curves show cases in which the dUrca sets in within the CSE branch.
We write in each plot the results for the maximum mass and the fiducial radius of a $1.4M_\odot$ star, $R_{1.4}$, with (without) the dUrca constraint in black (blue). The left plots show that the dUrca constraint leads to a larger NS radius, therefore a stiffer EoS.
Including $\chi$EFT disfavors stiff EoSs, while astrophysical constraints are the most stringent and favor EoSs leading to smaller radii.
The combined posterior on the right panels shows that random (causal) $n_{\rm break}$ constrains the $R_{1.4}$ and $M_{\rm TOV}$ to $\approx 11.89$~km and $\approx 2.18$~M$_{\odot}$ ($\approx 11.91$~km and $\approx 2.27$~M$_{\odot}$).
Comparing the two treatments of the nucleonic metamodel branch, we find that extending the metamodel up to its causal limit (upper panels) favors larger NS masses and radii compared to randomly sampling $n_{\rm break}$ (bottom panels). 
While previous studies have pointed out that extending the nucleonic metamodel to high densities can bias EoS inference~\cite{Suleiman:2025dkn,Montefusco:2026jlq}, our results quantify the corresponding impact on global NS properties and demonstrate the sensitivity of the inferred $M_{\rm TOV}$ and $R_{1.4}$ to the assumed breakdown density.

Figure~\ref{fig:Combined_Random_nbreak_pub_mass_radius_hist_Skyrme} shows the mass-radius diagram for the nucleonic matter modeled by the extended-Skyrme functional. The general trend is equivalent to the metamodel case. In the upper panels, we note that Skyrme can probe more massive NS under the nucleonic assumption than the metamodel. This enables us to probe higher densities and to reduce noticeably the cases in which $n_{\rm DU}$ happens within the CSE branch. 
We also note that the Skyrme combined posterior with dUrca leads to $R_{1.4}\approx 12.2$~km and $M_{\rm TOV}\approx 2.16$~M$_{\odot}$, more or less regardless of the $n_{\rm break}$ construction. Compared to the metamodel, the Skyrme framework leads to a slightly larger canonical NS radius and a slightly lower TOV mass, although both are consistent within the given uncertainties. 
Generally, the extended-Skyrme mass-radius results agree well with the metamodel case. 

We show in Fig.~\ref{fig:distributions} the results for the tidal deformability of a 1.4~$M_{\odot}$ NS in the left panels, the nuclear empirical parameter $L_{\rm sym}$ in the center, and the proton fraction in the center of the canonical $1.4$~M$_{\odot}$ on the right panels. 
Again, the bottom panels show the case where we pick $n_{\rm break}$ randomly, and the top panels show the case where the nucleonic model is used up to its causal limit.
The priors are shown as dashed lines, where the dUrca (blue) prior tends toward higher $\Lambda_{1.4}$, $L_{\rm sym}$, and $Y_{p}$ than $\chi$EFT (green) and Astro (red). 
Solid lines show the posterior distributions for the different constraints. 
$\chi$EFT (green) and astro (red) are known to prefer softer EoS, and the obtained posteriors peak at lower values of $\Lambda_{1.4}$ and $L_{\rm sym}$.
The dUrca constraint provides a rather broad distribution for tidal deformability and $L_{\rm sym}$, showing negligible impact on NS macroscopic properties, as the $\chi$EFT and astrophysical constraints are substantially more informative for the inference.  
In contrast, the proton-fraction distribution on the right shows that the combined $\chi$EFT + Astro + dUrca posterior is concentrated around $Y_p\approx0.15$, indicating that dUrca provides the dominant constraint on the composition of a 1.4~$M_{\odot}$ NS. 
This contrasts with Ref.~\cite{Montefusco:2026dU}, where laboratory symmetry-energy data yield $Y_p \approx 0.07$--$0.09$ at $1.4\,M_\odot$ and disfavor dUrca onset below this mass. The difference is driven by the cooling data: there, dUrca is evaluated a posteriori within a single nucleonic metamodel, whereas here it enters the likelihood and may also set in beyond $n_{\rm break}$. We note that through the inclusion of Cassiopeia A, PSR B2334+61, and PSR B0833-45, based on the assumption that dUrca is active within these objects, it is to be expected that our analysis provides a higher proton fraction within the stars' center compared to~\cite{Montefusco:2026dU}. 
Without the dUrca likelihood, our $\chi$EFT + Astro posterior is compatible with theirs. If both results hold, rapid cooling near $1.4\,M_\odot$ then requires larger source masses or a non-nucleonic fast-cooling channel.

As previously, the use of the metamodel up to its causal limit leads to a moderate shift in the tidal deformability and the slope of the symmetry energy. The combined distribution (solid red) favors $\Lambda_{1.4}= 376_{-168}^{+209} (357_{-153}^{+191})$ for random (causal) $n_{\rm break}$, and $ L_{\rm sym} = 39.8_{-27.3}^{+44.9} ~\text{MeV}~ (52.7_{-37.5}^{+45.6} ~\text{MeV})$ for random (causal) $n_{\rm break}$, both for 95\% CI.
The distributions for the same quantities using the extended-Skyrme model for the nucleonic EoS are in good agreement with the metamodel case. The Skyrme posteriors show larger fractions of the nucleonic branch compared to the CSE branch than the metamodel. Constraints on the three quantities are tighter from the Skyrme EoS and also less sensitive to the choice of $n_{\rm break}$ compared to the metamodel,  
with $L_{\rm sym} = 43.7_{-34.6}^{+32.4} \text{MeV} (43.6_{-30.2}^{+29.9} \text{ MeV})$ for random (causal) $n_{\rm break}$, both for 95\% CI.

\begin{figure}
    \centering
    \includegraphics[width=1\linewidth]{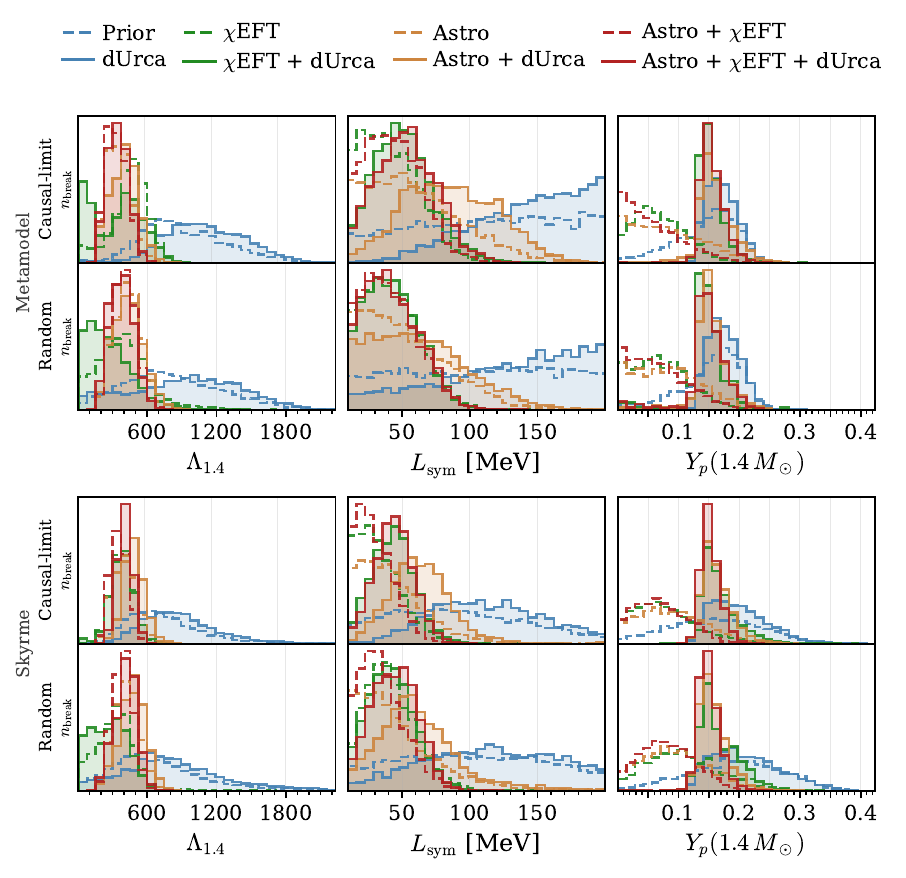}
    \caption{ Distributions of (left) tidal deformability for a canonical 1.4~$M_{\odot}$, (center) the slope of the symmetry energy $L_{\rm sym}$, and (right) the proton fraction at the center of the canonical star with metamodel (top) and Skyrme (bottom) construction.
    Top (bottom) sub-panels on each EoS show the results for causal-limit (random) $n_{\rm break}$. Dashed (solid) lines show the prior (posterior) distributions. 
    We show the different constraints ($\chi$EFT, Astro, dUrca) using different colors.}
    \label{fig:distributions}
\end{figure}

\section{Conclusions} 
In this work, we have performed a multimessenger Bayesian analysis combining nuclear physics information from $\chi$EFT, macroscopic NS observables from gravitational waves and NICER observations, and constraints on rapid cooling through the dUrca process. 
To incorporate the latter consistently, we extended the \textsc{jester} framework to compute the composition of $\beta$-equilibrated matter self-consistently and determine the corresponding dUrca threshold. 
This provides a more complete assessment of the information available from multimessenger observations than analyses based solely on macroscopic EoS observables, since dUrca constraints require explicit information about the microscopic composition, in particular the proton fraction $Y_p$. 
Within the assumptions adopted here for the nucleonic models and the dUrca likelihood, we find that the dUrca constraint has only a minor impact on macroscopic NS properties such as the tidal deformability or radius, while additional information on the composition can be provided. This study represents a step towards incorporating composition information systematically into multimessenger EoS inference. 

Notably, the qualitative results remain consistent across the four EoS constructions: metamodel and extended-Skyrme, each with random or causal breakdown of the nucleonic hypothesis. 
This consistency matters for the  composition results, since $Y_p$ at a given mass is only defined where the
nucleonic description still holds. 
In this respect, the extended-Skyrme functional retains the nucleonic hypothesis up to higher densities and NS masses than the metamodel, and is correspondingly less sensitive to the $n_{\rm break}$ construction, so that a larger fraction of the posterior provides a well-defined composition for canonical-mass stars. 
We stress that this concerns the availability of a nucleonic composition rather than its value, which the cooling constraint sets as discussed above.

The constraint of Eq.~\eqref{eq:likelihooddurca} enters entirely through the comparison between the measured mass of a cooling object and the threshold mass $M_{\rm DU}$, yet for most rapidly cooling sources the mass is itself derived from the thermal modeling we wish to interpret. Independent and complementary determinations of mass, surface temperature and age for the same object would remove this circularity~\cite{Beloin:2018fyp,Beznogov:2015qra}, with low-mass NS carrying particular weight, because they are not expected to reach $Y_{\rm DU}$ for any EoS compatible with current nuclear physics constraints~\cite{Yakovlev:2004iq, Margueron:2017lup}.

\section{Acknowledgements}
\label{section:acknowledgements}
M.~D.~D.\ acknowledges support from a doctoral research grant from the German Academic Exchange Service (DAAD).
T.W.\ is supported by the research program of the Netherlands Organization for Scientific Research (NWO) under grant number OCENW.XL21.XL21.038.
T.~D.\ acknowledges funding from the European Union (ERC, SMArt, 101076369). Views and opinions expressed are those of the authors only and do not necessarily reflect those of the European Union or the European Research Council. Neither the European Union nor the granting authority can be held responsible for them. 
The simulations were performed on DFG-funded research cluster Jarvis at the University of Potsdam (INST 336/173-1; project number: 502227537).

\bibliography{main.bib}

\end{document}